\documentclass[twocolumn,seceq]{jpsj2}

\title{Hyperbolic Deformation on Quantum Lattice Hamiltonians}

\author
{Hiroshi {\sc Ueda}$^1$ and Tomotoshi {\sc Nishino}$^2$ }

\inst
{
$^{1)}$ Department of Material Engineering Science, 
Graduate School of Engineering Science, Osaka University, Osaka 560-8531\\
$^{2)}$ Department of Physics, Graduate School of Science, Kobe University, Kobe 657-8501
}

\recdate{\today}

\abst
{
A group of non-uniform quantum lattice Hamiltonians in one dimension is introduced,
which is related to the hyperbolic $1 + 1$-dimensional space.
The Hamiltonians contain only nearest neighbor interactions whose strength is proportional 
to $\cosh j \lambda$, where $j$ is the lattice index and where $\lambda \ge 0$ is a 
deformation parameter. In the limit $\lambda \rightarrow 0$ the
Hamiltonians become uniform.  Spacial translation of the deformed Hamiltonians is
induced by the corner Hamiltonians. As a simple example, we investigate
the ground state of the deformed $S = 1/2$ 
Heisenberg spin chain by use of the 
density matrix renormalization group (DMRG) method. It is shown that 
the ground state is dimerized when $\lambda$ is finite.
Spin correlation function show exponential decay, and the
boundary effect decreases with increasing $\lambda$. 
}

\kword{DMRG, Corner Hamiltonian, Regularization, Renormalization Grup, }

\begin{document}
\sloppy
\maketitle

\section{Introduction}

The density matrix renormalization group (DMRG) method has been applied to 
various problems in low dimensional correlated physics.~\cite{White1,White2,Peschel,Schollwoeck}
The method contains the block spin transformation in its 
formulation, but the relation with Wilson's renormalization group~\cite{Wilson1} 
(RG) is not clear, since the hierarchy in energy scale is missing in the uniform lattice
Hamiltonians. Recently Okunishi considered a group of half-infinite non-uniform systems, 
whose lattice Hamiltonians 
\begin{equation}
H( \lambda ) 
= \sum_{j=1}^{N} \,\, \Lambda^j_{~} \,\, h_{j, j+1}^{~}
= \sum_{j=1}^{N} \,\, e^{j \lambda}_{~} \,\, h_{j, j+1}^{~} 
\end{equation}
are sum of neighboring interaction terms $h_{j, j+1}^{~}$, 
where $j$ is the lattice index and where $N+1$ is the system size.~\cite{Oku1}
He applied numerical RG method to this system, where the application
is a one-parameter deformation --- the {\it exponential deformation} --- 
to the real space RG scheme introduced by Xiang.~\cite{Xiang1,Xiang2}
It is shown that the energy scale introduced by the 
deformation parameter $\Lambda = e^{\lambda}_{~} \ge 1$ regularize the distribution
of low energy excitations, even for the models that are gapless 
when $\lambda = 0$ in the large $N$ limit.

It is possible to introduce Okunishi's deformation scheme to the 
DMRG method for the purpose of regularizing low energy excitations.
A primitive way is to joint the deformed half-infinite Hamiltonians in Eq.~(1.1) 
at the origin
\begin{equation}
H^{\rm joint}_{~}( \lambda) = \sum_{j=-N}^{N} \,\, e^{| j | \lambda}_{~} \,\, h_{j, j+1}^{~} 
\end{equation}
to construct the whole system of size $2N + 2$.
The interaction strength increases with $| j |$ toward the boundary of the
system, in contrast to the smooth boundary condition proposed by 
Veki\'c and White.~\cite{White3}
The construction of the above Hamiltonian is, however, rather 
artificial in the point that one may choose 
arbitrary increasing function of $| j |$ instead of $e^{| j | \lambda}_{~}$. In this 
article we introduce a natural candidate
\begin{eqnarray}
H^{\cosh}_{~}( \lambda ) 
\!\!\! &=& \!\!\!
\frac{1}{2} \sum_{j=-N}^{N} \! e^{ j  \lambda}_{~} \, h_{j, j+1}^{~} + 
\frac{1}{2} \sum_{j=-N}^{N} \!e^{ - j  \lambda}_{~} \, h_{j, j+1}^{~} \nonumber\\
\!\!\! &=& \!\!\!
\sum_{j=-N}^{N} \,\, \cosh( j  \lambda ) \,\, h_{j, j+1}^{~} \, ,
\end{eqnarray}
which keeps some aspects of translational invariance even when 
$\lambda > 0$. It is possible
to find a geometric interpretation of $H^{\cosh}_{~}( \lambda )$ as a 
time boost in the hyperbolic $1 + 1$-dimensional space, as discussed
at the end of this article.

In the next section we shortly review the exponential deformation on the 
lattice Hamiltonian. Eigenvalue distribution is considered in the 
large system size limit. 
In \S 3 we explain the details of the deformed Hamiltonian 
$H^{\cosh}{~}( \lambda )$ in Eq.~(1.3). 
It is shown that the lattice translation is related to the deformed corner Hamiltonian.
Recursive construction of  $H^{\cosh}{~}( \lambda )$ is also considered. 
Ground state property of the deformed system is analyzed in \S 4
in the case of the $S = 1/2$  
Heisenberg spin chain. It is shown that the spin
correlation function of the ground state decays exponentially when $\lambda$ is 
finite, and the state is dimerized. The effect of deformation on the quantum 
entropy is observed. 
Conclusions are summarized in the last section. Quantum-classical 
correspondence of the deformed Hamiltonians and several conjectures
are discussed.

\section{Exponential Deformation}

Consider a group of 1D quantum Hamiltonians 
\begin{equation}
H = \sum_{j = -N}^{N} h_{j, j+1}^{~} 
\end{equation}
on the lattice of size $2N + 2$,
where $h_{j, j+1}^{~}$ represents the interaction between neighboring sites 
labeled by $j$ and $j+1$. A typical example is the antiferromagnetic 
Heisenberg spin Hamiltonian
\begin{equation}
H_{H}^{~} = J \sum_{j = -N}^{N} 
{\bf s}_j^{~} \cdot {\bf s}_{j+1}^{~}  \, ,
\end{equation}
where $J \ge 0$ represents the interaction parameter. 
Before considering the deformed Hamiltonian 
$H^{\cosh}_{~}( \lambda )$ in Eq.~(1.3), let us observe effects of the 
exponential deformation in Eq.~(1.1). For latter convenience 
we treat the system whose linear size is $2N + 2$. The 
exponentially deformed Hamiltonian  is then written as
\begin{equation}
H^{\rm exp}_{~}( \lambda ) = \sum_{j = -N}^{N} e^{j \lambda}_{~} \, h_{j, j+1}^{~} \, ,
\end{equation}
where the deformation parameter $\lambda$ is real and positive.~\cite{complex} 
When $\lambda = 0$ the above Hamiltonian $H^{\rm exp}_{~}( \lambda )$ coincides 
with the uniform Hamiltonian $H$ in Eq.~(2.1).

It is known that the factor $\Lambda = e^\lambda_{~}$ controls the
eigenvalue structure.~\cite{Wilson1,Oku1} In order to observe the fact briefly, 
let us consider the infinite system size limit $N \rightarrow \infty$. 
To simplify the discussion we assume that the ground state energy $E_0^{~}$ is zero,
and all other eigenvalues are positive. This assumption can be satisfied by adding appropriate
constant to each neighboring interaction $h_{j, j+1}^{~}$.~\cite{math} 

Consider a right shift operation $S$ that moves
the lattice sites by one to the right direction. 
It is obvious that $S^\dagger_{~}$, the conjugate of $S$, represents
the left shift operation, and therefore $S S^\dagger_{~} = S^\dagger_{~} S = 1$ is
satisfied. If we apply $S$ to $H^{\rm exp}_{~}( \lambda )$ when the system size is
infinite, we obtain the following relation
\begin{eqnarray}
S \, H^{\rm exp}_{~}( \lambda ) \, S^\dagger_{~} \!\!\! &=&  \!\!\!
\sum_{j = -\infty}^{\infty} e^{j \lambda}_{~} \, 
\left( S \, h_{j, j+1}^{~} \, S^\dagger_{~} \right) \\
\!\!\! &=&  \!\!\!
\sum_{j = - \infty}^{\infty} e^{j \lambda}_{~} \, h_{j+1, j+2}^{~} \nonumber\\
\!\!\! &=&  \!\!\!
\sum_{j = - \infty}^{\infty} e^{(j-1) \lambda}_{~} \, h_{j, j+1}^{~} = 
e^{-\lambda}_{~} \, H^{\rm exp}_{~}( \lambda ) \, . \nonumber
\end{eqnarray}
As a result of translation the deformation parameter $e^{j \lambda}_{~}$ is
modified to $e^{( j - 1 ) \lambda}_{~}$, and this modification can simply
be expressed by multiplying the factor $e^{- \lambda}_{~}$ to 
$H^{\rm exp}_{~}( \lambda )$. 
This translation property in $H^{\rm exp}_{~}( \lambda )$ restricts the 
eigenvalue structure, which is obtained from the eigenvalue relation
\begin{equation}
H^{\rm exp}_{~}( \lambda ) \, | \Psi \rangle = E \, | \Psi \rangle \, .
\end{equation}
If there is an eigenstate $| \Psi \rangle$ the shifted state $S \, | \Psi \rangle$ is
also an eigenstate, since we have the relation
\begin{equation}
\left[ S \, H^{\rm exp}_{~}( \lambda ) \, S^\dagger_{~} \right] S \, | \Psi \rangle 
= S \, H^{\rm exp}_{~}( \lambda ) \, | \Psi \rangle
= E S \, | \Psi \rangle \, ,
\end{equation}
and using the relation in Eq.~(2.4) we can verify that
\begin{equation}
\left[ e^{- \lambda}_{~} \, H^{\rm exp}_{~}( \lambda ) \right] \, S \, | \Psi \rangle 
= E \, S \, | \Psi \rangle 
\end{equation}
is satisfied.
Thus if the eigenvalue $E$ in Eq.~(2.5) is positive, 
there is a family of eigenvalues 
\begin{equation}
\ldots, \, e^{-2\lambda}_{~} \, E, e^{-\lambda}_{~} \, E,
\, E, \, e^{\lambda}_{~} \, E, \, 
e^{2 \lambda}_{~} \, E, \, \ldots \,\, ,
\end{equation}
that are equidistant in logarithmic scale. Such a positive energy
eigenstate $| \Psi \rangle$ is not translationally invariant, and
the orthogonality
\begin{equation}
\langle \Psi | \, S \, | \Psi \rangle = 0 
\end{equation}
is satisfied. 

It should be noted that presence of periodic eigenstates are not excluded. 
For example, if there is unique zero-energy eigenstate $| \Phi \rangle$, 
it is translationally invariant. This is because Eq.~(2.7) shows that $S \, | \Phi \rangle$ is
also the zero energy state. Thus we can say that if the zero-energy state is
unique, it satisfies the translational invariance
\begin{equation}
S \, | \Phi \rangle = | \Phi \rangle \, .
\end{equation}
As an extension one can consider 
digenerated case, where there are two zero-energy eigenstates
 $| \Phi_a^{~} \rangle$ and $| \Phi_b^{~} \rangle$ that satisfies
\begin{eqnarray}
| \Phi_b^{~} \rangle \!\!\! &=& \!\!\! S \, | \Phi_a^{~} \rangle \nonumber\\
| \Phi_a^{~} \rangle \!\!\! &=& \!\!\! S \, | \Phi_b^{~} \rangle \, .
\end{eqnarray}
This is the case when there is dimerization in the ground state.
This degeneracy would be lifted by the effect of boundary when the system
size $2N + 2$ is finite.
It is straightforward to extend the argument of degeneracy to trimerized state, etc.

It is possible to consider various generalizations of $H^{\rm exp}_{~}( \lambda )$.
As an example one can consider the deformed tight-binding Hamiltonian
\begin{eqnarray}
H^{\rm exp}_{\rm t.b.}( \lambda ) 
\!\!\! &=& \!\!\! \sum_{j=-\infty}^{\infty} e^{j \lambda}_{~}
\biggl[
- t \, ( c_{j+1}^\dagger c_j^{~} + c_j^\dagger c_{j+1}^{~} )  \nonumber\\
\!\!\! &~& \!\!\! 
+ \, (-1)^j_{~} \, \frac{\Delta}{2} \, 
( c_j^\dagger c_j^{~} - c_{j+1}^\dagger c_{j+1}^{~} )
\biggr]
\end{eqnarray}
for spinless lattice Fermions,
where $t$ represents the hopping parameter and where $\Delta$ the band gap. 
Since this Hamiltonian contains oscillating potential, the translation period 
is 2-site when $\lambda = 0$. Thus for this deformed Hamiltonian 
$H^{\rm exp}_{\rm t.b.}( \lambda )$ 
one should modify the relation Eq.~(2.4) according to this period. It can be
verified that all the one-particle states $| \Psi \rangle$ satisfy 
the orthogonality in Eq.~(2.9), and are represented by localized wave
functions similar to wavelet basis function. 
The half-filled state $| \Phi \rangle$ has finite 
excitation gap, where $| \Phi \rangle$ is periodic and 
satisfies $S^2_{~} \, | \Phi \rangle = | \Phi \rangle$.
When $\lambda = 0$ the one-particle eigenfunctions and energy spectrum is 
explained by the Bloch's theorem. It is not trivial how such an 
energy structure is destructed by the introduction of exponential deformation.
It is straightforward to generalize the exponential deformation to systems
that contain interactions of longer range.

\section{Hyperbolic Deformation}

The eigenvalue distribution of $H^{\rm exp}_{~}( \lambda )$ explained in the 
last section prevents numerical study of the {\it bulk property} of the 
system around the center $j = 0$.
This is because the energy scale in the left side of the system ($j < 0$) is smaller 
than that at the center, and to apply the DMRG method to such system is difficult.
This problem can be avoided if we take an average between 
$H^{\rm exp}_{~}( \lambda )$ and $H^{\rm exp}_{~}( - \lambda )$ as
\begin{eqnarray}
H^{\cosh}_{~}( \lambda ) 
&=& \frac{1}{2} 
\bigl[ H^{\rm exp}_{~}( \lambda ) + H^{\rm exp}_{~}( - \lambda ) \bigr] \nonumber\\
&=&
\sum_{j=-N}^{N}  \cosh j \lambda  \,\, h_{j, j+1}^{~} \, .
\end{eqnarray}
We call the deformation from $H$ in Eq.~(2.1) to $H^{\cosh}_{~}( \lambda )$ 
introduced here as the {\it hyperbolic deformation} in the following. 

Let us extend the shift operation $S$ and its conjugate $S^{\dagger}_{~}$ to 
Hamiltonians of finite size systems. A natural way is to consider that the operation
modifies the coefficients of the neighboring interactions as follows
\begin{equation}
\cosh j \lambda \rightarrow \cosh (j-1) \lambda \, .
\end{equation}
Then the shift operation on $H^{\cosh}_{~}( \lambda )$ is defined as
\begin{equation}
S \, H^{\cosh}_{~}( \lambda ) \, S^{\dagger}_{~} =
\sum_{j=-N}^{N}  \cosh (j-1) \lambda  \,\, h_{j, j+1}^{~} \, .
\end{equation}
Taking the weighted difference between $H^{\cosh}_{~}( \lambda )$ 
and $S \, H^{\cosh}_{~}( \lambda ) \, S^{\dagger}_{~}$ we obtain
the relation
\begin{eqnarray}
&~& \!\!\!\!\! H^{\cosh}_{~}( \lambda ) - \frac{1}{\cosh \lambda} \, 
S \, H^{\cosh}_{~}( \lambda ) \, S^{\dagger}_{~} \\
&=& \!\!\!\!\! \tanh \lambda \, \sum_{j=-N}^{N} 
\sinh j \lambda \,\, h_{j, j+1}^{~} 
= \tanh \lambda \, H^{\sinh}_{~}( \lambda ) \nonumber \, ,
\end{eqnarray}
where $H^{\sinh}_{~}( \lambda )$ introduced here represents deformed
Hamiltonian of another type
\begin{equation}
H^{\sinh}_{~}( \lambda ) = \sum_{j=-N}^{N} \sinh j \lambda \,\, h_{j, j+1}^{~} \, ,
\end{equation}
which is decoupled at the origin $j = 0$. Similar to Eq.~(3.4),
the deformed Hamiltonian $H^{\cosh}_{~}( \lambda )$ can be
obtained from $H^{\sinh}_{~}( \lambda )$ by the following weighted difference
\begin{equation}
H^{\sinh}_{~}( \lambda ) - \frac{1}{\cosh \lambda} \, 
S \, H^{\sinh}_{~}( \lambda ) \, S^{\dagger}_{~} = \tanh \lambda \, 
H^{\cosh}_{~}( \lambda ) \, .
\end{equation}
The relations Eqs.~(3.4) and (3.6) can be regarded as one parameter
deformation to the translational invariance $S H S^{\dagger}_{~} = H$,
which is satisfied by the uniform Hamiltonian in Eq.~(2.1).

Following the convention in the infinite system DMRG method, 
let us divide $H^{\cosh}_{~}( \lambda )$ into three parts
\begin{equation}
H^{\cosh}_{~}( \lambda ) = H_L^{~}( \lambda ) + h_{0, 1}^{~} + H_R^{~}( \lambda ) \, ,
\end{equation}
where $H_L^{~}( \lambda )$ and $H_R^{~}( \lambda )$ are defined as follows
\begin{eqnarray}
H_L^{~}( \lambda ) \!\!\! &=& \!\!\! \sum_{j=-N}^{-1}  \cosh j \lambda  \,\, h_{j, j+1}^{~}
\nonumber\\
H_R^{~}( \lambda ) \!\!\! &=& \!\!\! \sum_{j=1}^{N}  \cosh j \lambda  \,\, h_{j, j+1}^{~} \, .
\end{eqnarray}
We also divide $H^{\sinh}_{~}( \lambda )$ in the same manner
\begin{equation}
H^{\sinh}_{~}( \lambda ) \,\,\, = \,\,\, C_L^{~}( \lambda ) + C_R^{~}( \lambda ) \, ,
\end{equation}
where $C_{\rm L}^{~}( \lambda )$ and $C_{\rm R}^{~}( \lambda )$ are 
defined as follows
\begin{eqnarray}
C_L^{~}( \lambda ) \!\!\! &=& \!\!\! 
\sum_{j = -N}^{-1} \sinh j \lambda \,\, h_{j, j+1}^{~} \nonumber\\
C_R^{~}( \lambda ) \!\!\! &=& \!\!\! 
\sum_{j = 1}^{N} \sinh j \lambda \,\, h_{j, j+1}^{~} \, .
\end{eqnarray}
These are deformations to the corner Hamiltonian,~\cite{Baxter,Oku2} 
since in the limit $\lambda \rightarrow 0$ we obtain the relation
\begin{equation}
\lim_{\lambda \rightarrow 0} \frac{C_R^{~}( \lambda )}{\sinh \lambda} = 
\sum_{j=1}^N j \, h_{j, j+1}^{~} \, .
\end{equation}

We have shown the relation between $H^{\cosh}_{~}( \lambda )$ and 
$H^{\sinh}_{~}( \lambda )$ for the same system size $2N + 2$. We then focus
on recursion relations, which connects systems of different sizes.
Let us introduce Baxter's star notation~\cite{Baxter,Oku2}
\begin{eqnarray}
H_R^{*}( \lambda ) \!\!\! &=& \!\!\! \sum_{j = 2}^{N} \cosh (j-1) \lambda \,\, h_{j, j+1}^{~} 
\nonumber\\
C_R^{*}( \lambda ) \!\!\! &=& \!\!\! \sum_{j = 2}^{N} \sinh (j-1) \lambda \,\, h_{j, j+1}^{~} \, .\end{eqnarray}
We then obtain recursion relation
\begin{eqnarray}
C_R^{~}( \lambda ) 
\!\!\! &=& \!\!\!
\sum_{j = 1}^{N} \sinh \bigl[ (j-1) \lambda + \lambda \bigr]  \,\, h_{j, j+1}^{~} \\
\!\!\! &=& \!\!\!
\cosh\lambda \,\, C_R^{*}( \lambda ) + 
\sinh\lambda \, \bigl[ h_{1, 2}^{~} +  H_R^{*}( \lambda ) \bigr] \, ,
\nonumber
\end{eqnarray}
and similarly we obtain
\begin{eqnarray}
H_R^{~}( \lambda ) \!\!\! &=& \!\!\!
\sum_{j = 1}^{N} \cosh \bigl[ (j-1) \lambda + \lambda \bigr]  \,\, h_{j, j+1}^{~} \\
\!\!\! &=& \!\!\!
\cosh\lambda \, \bigl[ h_{1, 2}^{~} + H_R^{*}( \lambda ) \bigr] + 
\sinh\lambda \,\, C_R^{*}( \lambda ) \, .
\nonumber
\end{eqnarray}
If we introduce the double star notations
\begin{eqnarray}
H_R^{**}( \lambda ) \!\!\! &=& \!\!\! \sum_{j = 3}^{N} \cosh (j-2) \lambda \,\, h_{j, j+1}^{~} 
\nonumber\\
C_R^{**}( \lambda ) \!\!\! &=& \!\!\! \sum_{j = 3}^{N} \sinh (j-2) \lambda \,\, h_{j, j+1}^{~}
\, ,
\end{eqnarray}
we can decouple the recursion relations as follows
\begin{eqnarray}
H_R^{~}( \lambda ) \!\!\! &=& \!\!\!
\cosh \lambda \,\, h_{1, 2}^{~} - h_{2, 3}^{~} + 
2 \cosh \lambda \,\, H_R^{*}( \lambda ) - H_R^{**}( \lambda )
\nonumber\\
C_R^{~}( \lambda ) \!\!\! &=& \!\!\!
\sinh \lambda \,\, h_{1, 2}^{~} + 
2 \cosh \lambda \,\, C_R^{*}( \lambda ) - C_R^{**}( \lambda ) \, .
\end{eqnarray}
These relations would be of use when one applies numerical
renormalization group methods~\cite{White1,White2,Oku1,Oku2} to 
the deformed Hamiltonian $H^{\cosh}_{~}( \lambda )$ in order to
obtain its eigenstates.

\section{Numerical Observations}

One might conjecture that the hyperbolic deformation violates uniform property of the
system, since the bond interaction strength is modified. But for the ground state
this intuition is not always true. For example, one can show that the 
valence bond solid (VBS) state of $S = 1$ spin chains is not violated by 
the hyperbolic (or even exponential) deformation. We observe another example,
the ground state of the deformed $S = 1/2$ Heisenberg spin chain in this section.

\begin{figure}[h]
\begin{center}
\includegraphics[width=80mm]{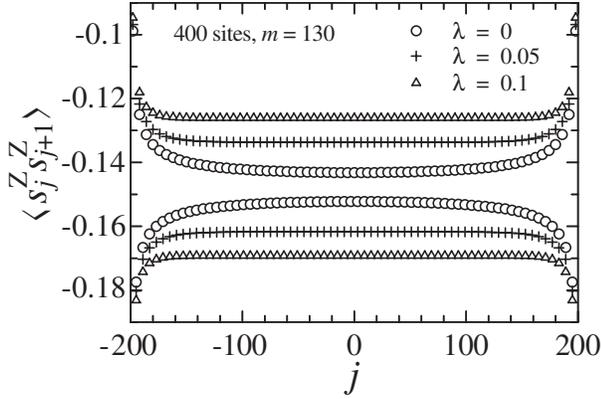}
\end{center}
\caption{\label{fig:2} Nearest neighbor spin correlation function 
$\langle {s}_j^{\rm Z} \,  {s}_{j+1}^{\rm Z} \rangle$ of 
the deformed $S = 1/2$ Heisenberg model. In all cases shown here the
function contains even-odd oscillation, which decays very slowly only 
when $\lambda = 0$.}
\end{figure}

Figure 1 shows the nearest neighbor spin correlation function
$\langle {s}_j^{\rm Z} \, {s}_{j+1}^{\rm Z} \rangle$ calculated for the ground state
of $400$-site system when $\lambda = 0$, $0.05$, and $0.1$.
We keep $m = 130$ states at most for the block spin variables in the
calculation by the finite system DMRG method. When $\lambda = 0$ the 
correlation function show even-odd oscillation with respect to $j$, and the 
oscillation slowly decays from the boundary to the center of the system.
It is known that the decay is in power low, which 
represents the gapless nature of the undeformed $S = 1/2$ Heisenberg chain.
When $\lambda$ is finite, the oscillation is strongly stabilized, and the
boundary effect disappears rapidly. 

\begin{figure}[h]
\begin{center}
\includegraphics[width=80mm]{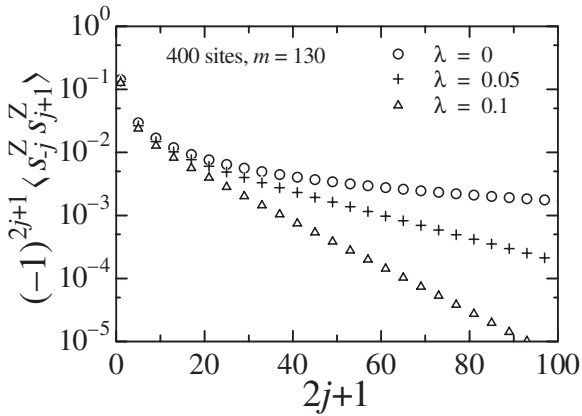}
\end{center}
\caption{\label{fig:2} Decay of the correlation function 
$(-1)^{2j+1}_{~} \, \langle {s}_{-j}^{\rm Z} \, {s}_{j+1}^{\rm Z} \rangle$ with respect to the
distance $2j + 1$.}
\end{figure}

Figure 2 shows the correlation function 
$| \, \langle {s}_{-j}^{\rm Z} \, {s}_{j+1}^{\rm Z} \rangle \, | 
= (-1)^{2j+1}_{~} \, \langle {s}_{-j}^{\rm Z} \, {s}_{j+1}^{\rm Z} \rangle$ with 
respect to the distance $2j+1$.
When $\lambda = 0.05$ and $\lambda = 0.1$ we observe exponential decay.
The correlation length $\xi$ obtained from the decay rate
is almost inverse proportional to  $\lambda$ as shown in Fig.~3, 
where $\xi \lambda \sim 0.134$ is satisfied. 
These calculated results suggest that the hyperbolic deformation enhances 
the local property of the system. To confirm this locality, we calculate the entanglement 
entropy. Figure 4 shows the bipartite entropy $S$ at the center of $400$-site system.
The value of $S$ decrease exponentially with $\lambda$, where 
$S = 1.145$ at the infinite $\lambda$ limit. 

\begin{figure}[h]
\begin{center}
\includegraphics[width=74mm]{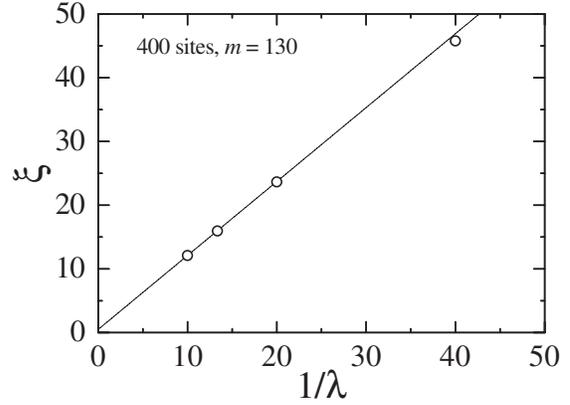}
\end{center}
\caption{\label{fig:3} Correlation length $\xi$ obtained from the
spin correlation function in Fig.~2.}
\end{figure}

\begin{figure}[h]
\begin{center}
\includegraphics[width=76mm]{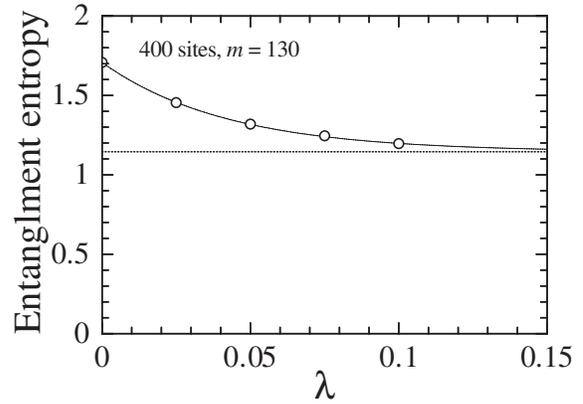}
\end{center}
\caption{\label{fig:4} Entanglement entropy $S$ as a function of $\lambda$.}
\end{figure}

\section{Conclusions and Discussions}

We have investigated the effect of hyperbolic deformation on 1D 
quantum lattice Hamiltonians. Numerical analysis on the deformed
$S = 1/2$ Heisenberg model  shows that the deformation introduces
dimerization in the ground state, and the local property is enhanced. 
The entangle entropy becomes finite even in the large system size
limit when $\lambda > 0$. 

Though the calculated system is one-dimensional it can possess
a dimerized ground state, because shift of dimerized pattern 
introduces macroscopic increase of energy expectation value; 
the dimer order might survive in finite temperature.
In this sense the deformed system has property of higher dimensional
systems. 

It would be interesting to consider whether the ground state is exactly 
represented by a matrix product state of finite matrix dimension
in the infinite $\lambda$ limit.
We conjecture that integer spin Heisenberg spin chains under strong
hyperbolic deformation have such finite dimensional matrix product ground states,
if appropriate boundary conditions are imposed.

The hyperbolic deformation can be used for scaling analysis of 
the ground state of undeformed system.
The two parameter scaling proposed by Tagliacozzo et. al~\cite{Tagliacozzo,Nishino}, where
the controllable parameters are the system size $N$ and the kept
number of block spin states $m$, can be modified to the scaling analysis
with respect to $\lambda$ and $m$. More simply, if one keeps sufficient
number of $m$ in numerical calculation, the finite size scaling with respect
to $N$ can be replaced by the finite $1/\lambda$ scaling.

Let us cast our eye to the quantum-classical correspondence.
If one considers the Trotter decomposition~\cite{Trotter,Suzuki} of the 
deformed Hamiltonian $H^{\cosh}_{~}( \lambda )$, one finds 
that the Hamiltonian describes real or imaginary time boost in the
hyperbolic $1+1$-dimensional space, which has constant negative curvature.
It is known that classical lattice models on the hyperbolic 2D space tend to show gaussian 
universality in their phase transition.~\cite{Rietman,Doyon,dAuriac,Shima,Hasegawa,
Nishino2,Nishino3,Ueda} This suggest
that if $H^{\cosh}_{~}( \lambda )$ describes quantum phase transition of
second order, it would subject to Gaussian universality class. 
Such a geometric interpretation may draw deformation of various type,
such as {\it spherical deformation} $H^{\sin}_{~}( \lambda )$.

The authors thank to K.~Okunishi for valuable discussions and critical reading
of this article.
This work is partially supported a Grant-in-Aid for Scientific Research from Japanese 
Ministry of Education, Culture, Sports, Science and Technology.


\begin{thebibliography}{99}
\bibitem{White1} S.R.~White: Phys. Rev. Lett. {\bf 69} (1992) 2863.
\bibitem{White2} S.R.~White: Phys. Rev. B {\bf 48} (1992) 10345.
\bibitem{Peschel} I.~Peschel, X.~Wang, M.~Kaulke, and K.~Hallberg (Eds.), 
{\it Density-Matrix Renormalization, A New Numerical Method in Physics}, 
Lecture Notes in Physics (Springer, Berlin 1999).
\bibitem{Schollwoeck} U.~Schollw\"{o}ck: Rev. Mod. Phys. {\bf 77} (2005) 259.
\bibitem{Wilson1} K.G.~Wilson: Rev. Mod. Phys. {\bf 47} (1975) 773.
\bibitem{Oku1} K.~Okunishi: J. Phys. Soc. Jpn. {\bf 76} (2007) 063001.
\bibitem{Xiang1} T.~Xiang and G.A.~Gehring, J. Magn. Magn. Matter. {\bf 104-107} (1992) 861.
\bibitem{Xiang2} T.~Xiang and G.A.~Gehring: Phys. Rev. B {\bf 48} (1993) 303.
\bibitem{White3} M.~Veki\'c and S.R.~White: Phys. Rev. Lett. {\bf 27} (1993) 4283.
\bibitem{complex} Since we consider hyperbolic geometry in the background of the
system, we restrict that the deformation parameter $\lambda$ is real. 
\bibitem{math} We don't argue if the deformed Hamiltonians are well defined 
mathematically in the infinite system size limit.
\bibitem{Baxter} R.J.~Baxter: {\it Exactly solved models in statistical mechanics} 
(Academic Press, London 1982).
\bibitem{Oku2} K.~Okunishi: J. Phys. Soc. Jpn. {\bf 74} (2005) 3186.
\bibitem{Tagliacozzo} L.~Tagliacozzo, T.R.~de Oliveria, S.~Iblisdir, and J.I.~Latorre:
Phys. Rev. B {\bf 78} (2008) 024410.
\bibitem{Nishino} T.~Nishino, K.~Okunishi, and M.~Kikuchi: J. Phys. Soc. Jpn. {\bf 65} 
(1996) 69.
\bibitem{Rietman} R.~Rietman, B.~Nienhuis,  and J.~Oitmaa: J. Phys. A: Math. Gen. 
{\bf 25}  (1992) 6577.
\bibitem{Doyon} B.~Doyon  and P.~Fonseca:  J. Stat. Mech. (2004) P07002.
\bibitem{dAuriac} J.C.~Angl\'es d'Auriac, R.~M\'elin, P.~Chandra,  and B.~Dou\c{c}ot:
J. Phys. A: Math. Gen. {\bf 34}  (2001)  675.
\bibitem{Shima} H.~Shima  and Y.~Sakaniwa: J. Phys. A: Math. Gen.   {\bf 39} (2006)  4921.
\bibitem{Hasegawa} I.~Hasegawa, Y.~Sakaniwa,  and H.~Shima:
Surf. Sci. {\bf 601}  (2007) 5232.
\bibitem{Nishino2} T.~Nishino:   J. Phys. Soc. Jpn. {\bf 65} (1996) 891.
\bibitem{Nishino3} T.~Nishino: J. Phys. Soc. Jpn. {\bf 66} (1997) 3040.
\bibitem{Ueda} K.~Ueda, R.~Krcmar, A.~Gendiar,  and T.~Nishino: J. Phys. Soc. Jpn.
 {\bf 76} (2007)  084004.
\bibitem{Trotter} H.~F.~Trotter: Proc. Am. Math. Soc. {\bf 10} (1959) 545.  
\bibitem{Suzuki} M.~Suzuki: Prog. Theor. Phys. {\bf 56} (1976) 1454.  
\end{thebibliography}
\end{document}